%% file: paper.tex
\documentclass[preprint,3p,times,twocolumn,sort&compress]{elsarticle}
\pdfoutput=1

\usepackage{graphicx}

\usepackage{preprintcover}

\PreprintCoverPaperTitle{Search for $W' \rightarrow t\bar{b}$ in the lepton plus jets final state in proton--proton collisions at a centre-of-mass energy of $\sqrt{s}$ = 8~TeV with the ATLAS detector}  

\PreprintIdNumber{CERN-PH-EP-2014-232}  

\PreprintCoverAbstract{A search for new charged massive gauge bosons, called $W'$,  is performed with the ATLAS detector at the LHC, in proton--proton collisions 
at a centre-of-mass energy of $\sqrt{s}$ = 8~\tev, using a dataset corresponding to an integrated luminosity of \lumi.
This analysis searches for $W'$ bosons  in the $W'  \rightarrow t\bar{b}$ decay channel in final states
with electrons or muons, using a multivariate method based on boosted decision trees. 
The search covers masses between 0.5 and 3.0~\tev, for right-handed or left-handed $W'$ bosons.
No significant deviation from the Standard Model expectation is observed and limits 
are set on the $W' \rightarrow t\bar{b}$ cross-section times branching ratio 
and on the $W'$-boson effective couplings as a function of the $W'$-boson mass using the CL$_s$ procedure.
For a left-handed (right-handed) $W'$ boson, masses below 1.70 (1.92)~\tev\ are excluded at 95\% confidence level.
}

\PreprintJournalName{Physics Letters B}

\usepackage{atlasphysics}
\usepackage{subfigure}
\usepackage{mathrsfs}
\usepackage{multirow}
\usepackage{extra_definitions}
\usepackage{url}
\usepackage{verbatim}
\usepackage{rotating}
\usepackage[debug]{hyperref}
\usepackage{placeins}

\def\ttbar {\mbox {$t\bar{t}$}}

\def\ifb {\mbox {$~{\rm fb}^{-1}$}}
\def\lumi{\mbox{$20.3\ifb$}}
\def\lumierr{\mbox{$20.3 \pm 0.6 \ifb$}}

\def\pt{\mbox{$p_{\text{T}}$}}

\def\MET{\ensuremath{E_{\mathrm{T}}^{\mathrm{miss}}}}
\def\kt{\mbox{$k_t$}}

\def\mtb{\mbox{$m_{t\bar{b}}$}}

\global\dimen\footins = 2cm 

\begin{document}

\begin{frontmatter}

\title{Search for $W' \rightarrow t\bar{b}$ in the lepton plus jets final state in proton--proton collisions at a centre-of-mass energy of $\sqrt{s}$ = 8~TeV with the ATLAS detector}

\begin{abstract}
A search for new charged massive gauge bosons, called $W'$,  is performed with the ATLAS detector at the LHC, in proton--proton collisions 
at a centre-of-mass energy of $\sqrt{s}$ = 8~\tev, using a dataset corresponding to an integrated luminosity of \lumi.
This analysis searches for $W'$ bosons  in the $W'  \rightarrow t\bar{b}$ decay channel in final states
with electrons or muons, using a multivariate method based on boosted decision trees. 
The search covers masses between 0.5 and 3.0~\tev, for right-handed or left-handed $W'$ bosons.
No significant deviation from the Standard Model expectation is observed and limits 
are set on the $W' \rightarrow t\bar{b}$ cross-section times branching ratio 
and on the $W'$-boson effective couplings as a function of the $W'$-boson mass using the CL$_s$ procedure.
For a left-handed (right-handed) $W'$ boson, masses below 1.70 (1.92)~\tev\ are excluded at 95\% confidence level.
\end{abstract}

\end{frontmatter}

\input{Introduction}

\input{Signal_Signature}

\input{Detector}

\input{Samples_Objects}

\input{Selection_Backgrounds}

\input{Analysis}

\input{Systematics}

\input{Result}

\input{Limit_Setting}

\FloatBarrier

\section{Summary}

This Letter describes a search for $W'~\rightarrow t\bar{b} ~\rightarrow~\ell \nu b\bar{b}$ in \lumi\ of proton--proton collisions
at a centre-of-mass energy of $\sqrt{s}$ = 8~\TeV\ with the ATLAS detector at the LHC.
Events with a lepton, missing transverse momentum and two $b$-tagged jets are selected. Multivariate discriminants are 
constructed using boosted decision trees.
By fitting these discriminants in data to the expectation, the consistency with the Standard Model 
background hypothesis is tested. 
The data are consistent with the Standard Model expectation and no evidence of $W'$-boson signal events is observed.
Exclusion limits at the 95$\%$ confidence level are set on the mass of the $W'$ boson and on its effective couplings.
Masses below 1.92 (1.80, 1.70)~\TeV\ are excluded for right-handed (left-handed without and with interference) $W'$ bosons, 
while the expected limit is 1.75 (1.57, 1.54)~\TeV.
The lowest observed (expected) limits on $g'/g$, obtained for a $W'$-boson mass of 0.75~\TeV, are 0.20 (0.28) and 0.16 (0.24) for 
left-handed and right-handed $W'$ bosons.

\input{Acknowledgements}

\bibliographystyle{atlasBibStyleWoTitle}

\bibliography{paper}

\clearpage
\onecolumn
\input{atlas_authlist} 

\end{document}

%% file: Introduction.tex
\section{Introduction}
\label{sec:into}
Many approaches to theories beyond the Standard Model (SM) introduce new charged vector currents mediated by heavy gauge bosons, usually called $W'$. 
For example, the $W'$ boson can appear in theories with universal extra dimensions, such as Kaluza--Klein excitations 
of the SM $W$ boson~\cite{Kaluza-Klein1,Kaluza-Klein2,Kaluza-Klein3},
or in theories that extend fundamental symmetries of the SM and propose a massive right-handed counterpart to the 
$W$ boson~\cite{left-right-model1,left-right-model2,left-right-model3}.
Little Higgs theories~\cite{little-Higgs} also predict a $W'$ boson.
The search for a $W'$ boson decaying to a top quark and a $b$-quark explores models
potentially inaccessible to searches for a $W'$ boson decaying into leptons~\cite{Wprime_lnu0,Wprime_lnu1,Wprime_lnu2,Wprime_lnu3}.
For instance, in the right-handed sector, the $W'$ boson cannot decay to a charged lepton and a right-handed 
neutrino if the latter has a mass greater than the $W'$-boson mass. 
Also, in several theories beyond the SM the $W'$ boson is expected to be coupled more strongly to the 
third generation of quarks than to the first and second generations~\cite{Muller:1996dj,StrongTop}.
Searches for a $W'$ boson decaying to the $t\bar{b}$ final state\footnote{
For simplicity, the notation ``$t\bar{b}$'' is used to describe both the $W'^{+} \to t\bar{b}$ and  $W'^{-} \to \bar{t}b$ processes.}
have been performed at the Tevatron~\cite{Abazov2011145,PhysRevLett.103.041801} in the leptonic top-quark decay channel
and at the Large Hadron Collider (LHC) in both the leptonic~\cite{ATLASWprime,CMSWprime,CMSWprime8TeV} and fully hadronic~\cite{ATLASWhad} final states, 
excluding right-handed $W'$ bosons with masses up to 2.05~\TeV\ at 95\% confidence level (CL).

%% file: Signal_Signature.tex
This Letter presents a search for $W'$ bosons using data collected in 2012 by the ATLAS detector~\cite{ATLASPaper} at the 
LHC, corresponding to an integrated luminosity of \lumi\
from proton--proton ($pp$) collisions at a centre-of-mass energy of 8~\TeV. 
The search is performed in the $W' \rightarrow t\bar{b} \rightarrow \ell \nu b\bar{b}$ decay channel,
where the lepton, $\ell$, is either an electron or a muon, using a multivariate method based on boosted decision trees. 
Right-handed and left-handed $W'$ bosons, denoted $W'_\text{R}$ and $W'_\text{L}$, respectively, are searched for in the mass range of 0.5 to 3.0~\tev.
A general Lorentz-invariant Lagrangian is used to describe the couplings of the $W'$ boson to fermions for various 
$W'$-boson masses~\cite{ZackSullivan,Duffty:2012rf}.
The mass of the right-handed neutrino is assumed to be larger than the mass of the $W'$ boson~\cite{ATLAS:2012ak}, 
thus allowing only hadronic decays of the $W'_\text{R}$ boson.
In the case of a $W'_\text{L}$ boson, leptonic decays are allowed and, since the signal
has the same event signature as SM $s$-channel single top-quark production,
an interference term between these two processes is taken into account~\cite{Interference}.

%% file: Detector.tex
\section{ATLAS detector}
\label{sec:detector}

Charged particles in the pseudorapidity\footnote{
ATLAS uses a right-handed coordinate system with its origin at the nominal interaction point in the centre of the detector 
and the $z$-axis along the beam pipe.  
The $x$-axis points from the interaction point to the centre of the LHC ring, and the $y$-axis points upward.  
Cylindrical coordinates ($r, \phi$) are used in the transverse plane, $\phi$ being the azimuthal angle around the beam pipe.  
The pseudorapidity is defined in terms of the polar angle $\theta$ as  $\eta$ = --ln tan($\theta/2$).
Observables labelled ``transverse'' are projected into the $x$--$y$ plane.
}
range $|\eta|$~$<$~$2.5$ are reconstructed with the inner detector, which consists of several layers of semiconductor detectors (pixel and microstrip),
and a straw-tube transition-radiation tracker, the latter covering $|\eta|$~$<$~$2.0$. 
The inner tracking detector system is immersed  in a homogeneous 2~T magnetic field provided by a superconducting solenoid.  
The solenoid is surrounded by a hermetic calorimeter that covers $|\eta|$~$<$~$4.9$ and provides three-dimensional reconstruction of particle showers.
The lead/liquid-argon electromagnetic compartment is finely segmented for $|\eta|$~$<$~$2.5$, where it plays an important role in electron identification.
Hadronic calorimetry is provided by a steel/scintillator-tiles calorimeter for $|\eta|$~$<$~$1.7$ and
by liquid-argon with copper or tungsten absorbers end-cap calorimeters that extend the coverage to $|\eta|=4.9$.
Outside the calorimeter, air-core toroids provide the magnetic field for the muon spectrometer.  Three stations of precision drift tubes and 
cathode-strip chambers provide an accurate measurement of the muon track curvature in the region $|\eta|$~$<$~$2.7$.  
Resistive-plate and thin-gap chambers provide muon triggering capability up to $|\eta|=2.4$.

%% file: Samples_Objects.tex
\section{Data and simulation samples}
\label{sec:samples}

The data used for this analysis were recorded using unprescaled single-electron and single-muon triggers. 
After stringent data-quality requirements, the amount of data corresponds to an integrated luminosity of \lumierr~\cite{lumiref}.

The $W'_\text{R}$ and $W_\text{L}'$ signals are generated with {\sc MadGraph5}~\cite{MadGraph5} using FeynRules~\cite{Degrande:2011ua,Alloul:2013bka}
and the {\sc CTEQ6L1}~\cite{cteq6} parton distribution function (PDF) set. {\sc Pythia8}~\cite{Pythia8} is used for parton showering and hadronisation.
Simulated samples are normalised to next-to-leading order (NLO) QCD calculations~\cite{Duffty:2012rf} using 
$K$-factors ranging from 1.15 to 1.35 depending on the mass and handedness of the $W'$ boson.
The model assumes that the $W'$-boson coupling strength to quarks, $g'$, is the same as for the $W$ boson:
$g'_\text{R} = 0$ and $g'_\text{L} = g$ ($g'_\text{R} = g$ and $g'_\text{L} = 0$) for left-handed (right-handed) $W'$ bosons, 
where $g$ is the SM SU(2)$_{\textrm{L}}$ coupling.
The total width of the left-handed (right-handed) $W'$ boson increases from 17 to 104~\gev\ (12 to 78~\gev) for masses between 0.5 and 3.0~\tev, 
where the decay to leptons is (is not) allowed~\cite{ZackSullivan}.
In order to account for the effect of the interference between $W'_\text{L}$-boson and $s$-channel single top-quark production 
dedicated  $pp \rightarrow W'_\text{L}/W \rightarrow t\bar{b} \rightarrow \ell\nu b\bar{b}$ samples are simulated,
using {\sc MadGraph5}, and assuming a destructive interference term~\cite{Interference}.
In addition, samples are generated for values of $g'/g$ up to 5.0, for several $W'$-boson (left- and right-handed) mass hypotheses.

Top-quark pair (\ttbar) and single top-quark $s$-channel and $Wt$ processes are simulated with the
{\sc Powheg}~\cite{powheg1, powheg2} generator, which uses a NLO QCD matrix element
with the CT10 PDFs~\cite{cteq6}. 
The parton shower and the underlying event are simulated using {\sc Pythia} v6.4~\cite{Pythia6}.
The $t$-channel single-top-quark process is modelled using the {\sc AcerMC} v3.8~\cite{acermc} generator
with the CTEQ6L1 PDFs and {\sc Pythia} v6.4.
The \ttbar\ cross-section is calculated at next-to-next-to-leading order (NNLO) in QCD including resummation of next-to-next-to-leading logarithmic 
soft gluon terms with {\sc top++2.0}~\cite{Cacciari:2011hy, Beneke:2011mq, Baernreuther:2012ws, Czakon:2013goa, Czakon:2012pz, Czakon:2012zr, Czakon:2011xx}.
The single top-quark cross-sections are obtained from approximate NNLO calculations~\cite{Kidonakis:2011wy,Kidonakis:2010tc,Kidonakis:2010ux}.
A top-quark mass of $172.5 \gev$ is assumed for the production of all simulated processes that include a top quark.

The {\sc Alpgen} leading-order multileg generator~\cite{SAMPLES-ALPGEN} with the CTEQ6L1 PDFs and {\sc Pythia} v6.4 
is used to generate vector bosons in association with jets: $W$+jets (including the contributions from 
$Wb\bar{b}$+jets, $Wc\bar{c}$+jets and $Wc$+jets) and $Z$+jets events.
Diboson samples ($WW$, $ZZ$, and $WZ$), where at least one of the bosons decays leptonically, are modelled using {\sc Herwig} v6.52~\cite{SAMPLES-HERWIG}
with the CTEQ6L1 PDFs.
The single-boson and diboson simulation samples are normalised to the production cross-sections calculated
at NNLO~\cite{Hamberg,FEWZ} and NLO~\cite{MCFMpaper} in QCD, respectively.

All generated samples are passed through a full simulation of the ATLAS detector~\cite{ATLASsim} based on GEANT4~\cite{SAMPLES-G4} 
and reconstructed using the same procedure as for collision data. 
Simulated events include the effect of multiple $pp$ collisions from the same and previous bunch-crossings
(in-time and out-of-time pileup) and are re-weighted to match the conditions of the data sample (20.7 interactions per bunch crossing on average).

%% file: Selection_Backgrounds.tex
\section{Object and event selections}
\label{sec:objects}

The search for $W'\rightarrow t\bar{b}$ events relies on the measurement of the following objects: 
electrons, muons, jets, and the missing transverse momentum. 
Electrons are identified as energy clusters in the electromagnetic calorimeter matched to reconstructed tracks 
in the inner detector~\cite{ele_calib,ele_recoid}.
Electron candidates are required to be isolated, using a fixed cone-size isolation criterion~\cite{ATLAS-CONF-2014-058}, from other objects in the event 
and from hadronic activity, to reduce the contamination from mis-reconstructed hadrons, electrons from heavy-flavour decays and photon conversions. 
Electrons are required to have transverse momentum, $\pt$, above 30~\gev\ and 
$\left| \eta \right | < 2.47$ with a veto on the barrel--endcap transition region 
in the range $1.37 < \left| \eta \right| < 1.52$. 

Muons are identified using the muon spectrometer and the inner detector~\cite{muon_perf}.  
A variable cone-size isolation criterion~\cite{Rehermann:2010vq,ATLAS-CONF-2014-058} is applied to reduce the contribution of muons from heavy-flavour decays.
Muon candidates are required to have \pt~$>30$~\gev\ and $\left| \eta \right| < 2.5$. 

Jets are reconstructed using the anti-\kt\ algorithm~\cite{Cacciari:2008gp} with a radius parameter $R=0.4$, 
using topological energy clusters as inputs~\cite{Clusters,JES}.
Jets are calibrated using energy- and $\eta$-dependent correction factors derived from simulation
and with residual corrections from in situ measurements~\cite{newJES}.
Jets are required to have \pt~$>25$~\gev\ and $|\eta|<2.5$.
To suppress jets from in-time pileup, at least 50\% of the scalar \pT~sum of the tracks associated with a jet 
is required to be from tracks associated with the primary vertex~\cite{JVF}. 
This requirement, called the ``jet vertex fraction'' requirement, is applied only for jets with \pt~$<50$~\gev\ and $|\eta|<2.4$.
The identification of jets originating from the hadronisation of $b$-quarks (``$b$-tagging'') is based on properties specific to $b$-hadrons, 
such as long lifetime and large mass.
This analysis uses a neural-network-based combination of several high-performance $b$-tagging algorithms~\cite{ATLAS-CONF-2011-102}.
The algorithm has an efficiency of 70\% (20\%, 0.7\%) for jets 
originating from $b$-quarks ($c$-quarks, light-quark/gluon) as obtained from simulated $t\bar{t}$ events.

The missing transverse momentum, \met, is the modulus of the vector sum of the transverse momentum in calorimeter cells associated with topological clusters,
and is further refined with object-level corrections from identified electrons, muons, and jets~\cite{ETmiss2010Paper,ATLAS-CONF-2012-101}.
This analysis requires events to have \met\ $> 35$~\gev\ to reduce the multijet background.

Candidate events are required to have exactly one lepton and two or three jets with exactly two of them identified as originating from a $b$-quark (denoted 2-tag events).
The multijet background contribution is further reduced by imposing a requirement on the sum of the $W$-boson transverse mass,\footnote{
Defined as \\ $m_{\mathrm{T}}(W)=\sqrt{(\pT(\ell)+\MET)^2-(p_x(\ell)+E_x^\mathrm{miss})^2-(p_y(\ell)+E_y^\mathrm{miss})^2}$,
where $E_x^\mathrm{miss}$ and $E_y^\mathrm{miss}$ are the $x$- and $y$-components of the \met~vector.} $m_{\mathrm{T}}(W)$, and $E_\mathrm{T}^{\mathrm{miss}}$: $m_{\mathrm{T}}(W) + E_\mathrm{T}^{\mathrm{miss}}>~60~\mathrm{\GeV}$. 
Events with exactly two (three) jets passing all the above selections define the 2-jet (3-jet) channel.

The signal region is defined by selecting events 
where the reconstructed invariant mass of the $t\bar{b}$ system, \mtb\ (see definition below), is larger than 330~\gev.
The acceptance times efficiency for the $W' \rightarrow t\bar{b}$ process in the lepton plus jets final state is 5.5\%, 2.2\% and 2.1\% (4.9\%, 2.2\%, 2.3\%) 
for a $W'_\text{R}$ ($W'_\text{L}$) boson with a mass of 1, 2 or 3~\tev, respectively. The drop in acceptance for high $W'$-boson masses
is due to the lepton failing the isolation criterion and to the decrease of the $b$-tagging efficiency.
A control region is defined by inverting the requirement on the $t\bar{b}$ invariant mass, \mtb~$<$~330~\gev, and is used to derive the normalisation of the $W$+jets background.

The method used to reconstruct the invariant mass of the $t\bar{b}$ system in the selected sample proceeds as follows.
The four-momentum of the top quark is reconstructed by adding the four-momenta of the $W$ boson
and of the $b$-tagged jet that gives the reconstructed invariant top-quark mass closest to the value used for generation (172.5~\gev).  
Thereafter, this $b$-tagged jet is called the ``top-jet" and is assumed to be the $b$-jet from the top quark.
In this calculation the transverse momentum of the neutrino is given by
the $x$- and $y$-components of the \met~vector, while the unmeasured 
$z$-component of the neutrino momentum is inferred by 
imposing a $W$-boson mass constraint on the lepton--neutrino system~\cite{tchpaper}.
The four-momentum of the $t\bar{b}$ system is then reconstructed by adding the four-momenta of the top quark to that of the remaining $b$-tagged jet.

\section{Background estimation}
\label{sec:bgestimation}

The \ttbar,  single-top-quark, diboson and $Z$+jets backgrounds are modelled using the simulation and are scaled to the theory predictions of the inclusive cross-sections.

The background originating from multijet events, where a jet is misidentified as a lepton or a non-prompt lepton appears isolated
(both referred to as a ``fake'' lepton), is estimated directly from data using the matrix method~\cite{ATLAS-CONF-2014-058}. 
The shape and normalisation of the multijet background are determined in both the electron and muon channels using this method.

The $W$+jets background is also modelled using the simulation, but in the case of the 2-jet channel the 
event yield for this process is derived from data to improve the modelling in this channel.
The number of $W$+jets events is estimated in the 2-jet control region as the number of data events observed after subtraction 
of all non-$W$+jets background sources described above.
This estimate is then extrapolated to the 2-jet signal region using the $W$+jets simulation.
For the 3-jet channel the $W$+jets background is scaled to the theory prediction.
%

%% file: Analysis.tex
\section{Analysis}
\label{sec:analysis}

The analysis strategy relies on a multivariate approach, based on the boosted decision tree (BDT) method 
using the framework of TMVA~\cite{ROOT-TMVA}, to enhance the separation between the 
signal and the background.
For each jet multiplicity and $W'$-boson handedness, a separate BDT is trained in the signal region. 
For the background, a mixture of top-quark, $W/Z$+jets, diboson and multijets samples, all weighted according to their relative abundances, is used.
The $W'$-boson sample used as signal in the BDT training and testing phases is chosen at a mass of 1.75~\tev\
since this gives the best expected exclusion limit on the $W'$-boson mass, compared to BDTs trained with other $W'$-boson mass samples.
This choice also ensures very good separation between the BDT shapes of signal and background for $W'$-boson masses of 1~\tev\ and above.
This analysis is thus sensitive to the presence of a signal over a wide mass range.

Ten (eleven) variables with significant separation power are identified in the 2-jet (3-jet) samples for the $W'$-boson search.
These are used as inputs to the BDTs.
The list of variables changes slightly depending on the chirality of the signal.

\begin{figure*}[!h!td]
\centering
\subfigure[]{\includegraphics[width=0.40\textwidth]{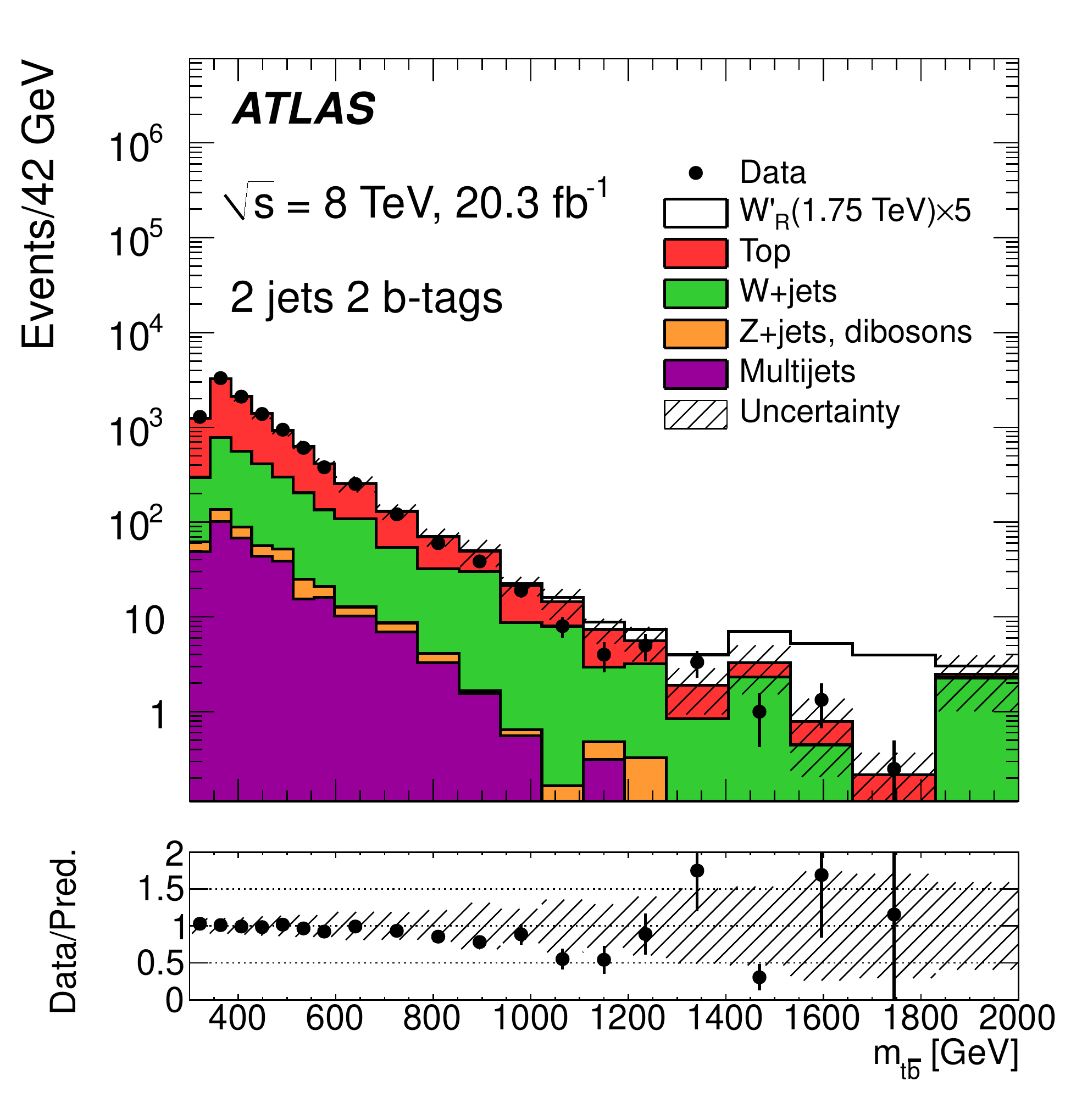}
\label{fig:mtb_SR_2jet}}
\subfigure[]{\includegraphics[width=0.40\textwidth]{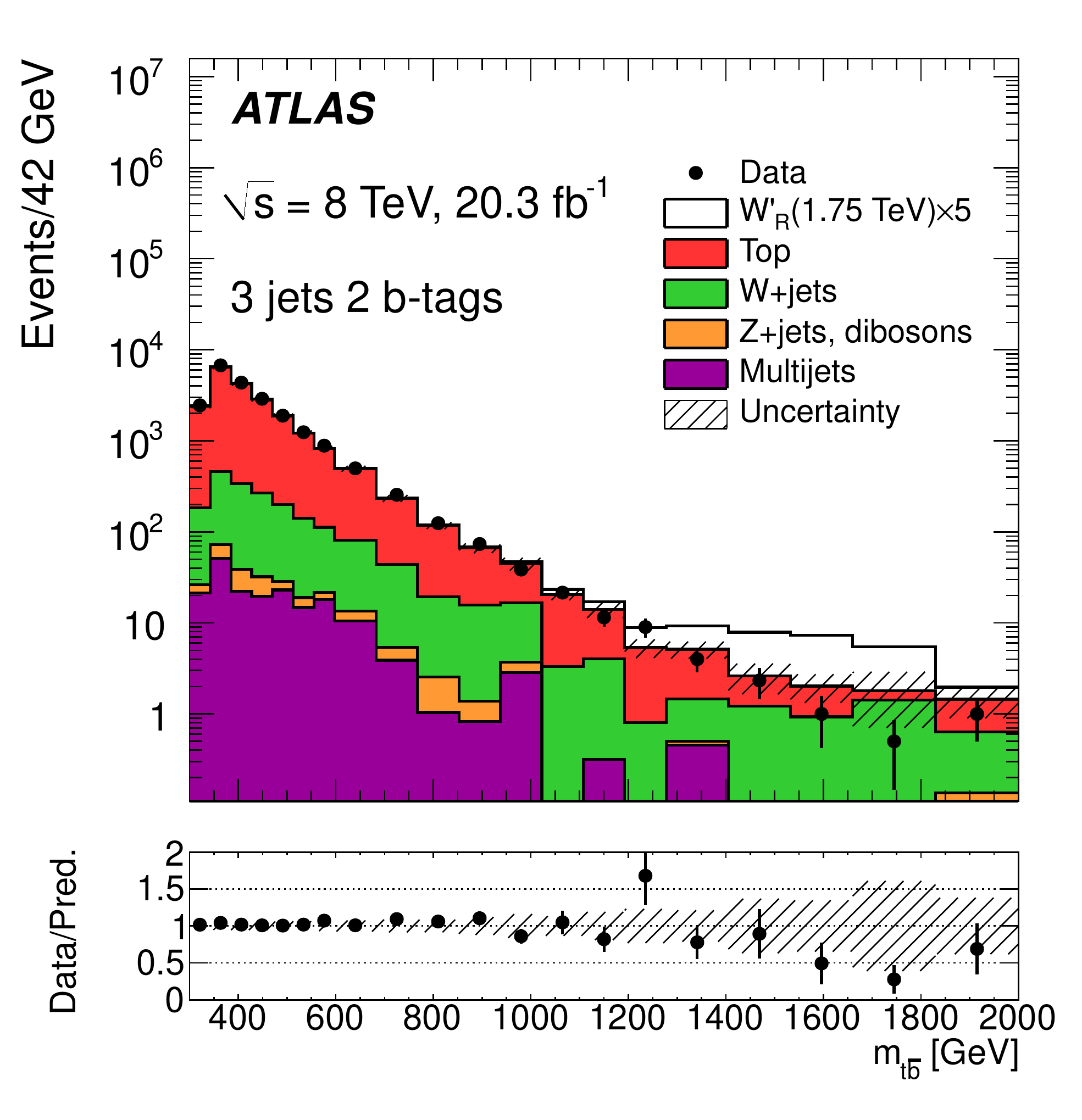}
\label{fig:mtb_SR_3jet}}
\subfigure[]{\includegraphics[width=0.40\textwidth]{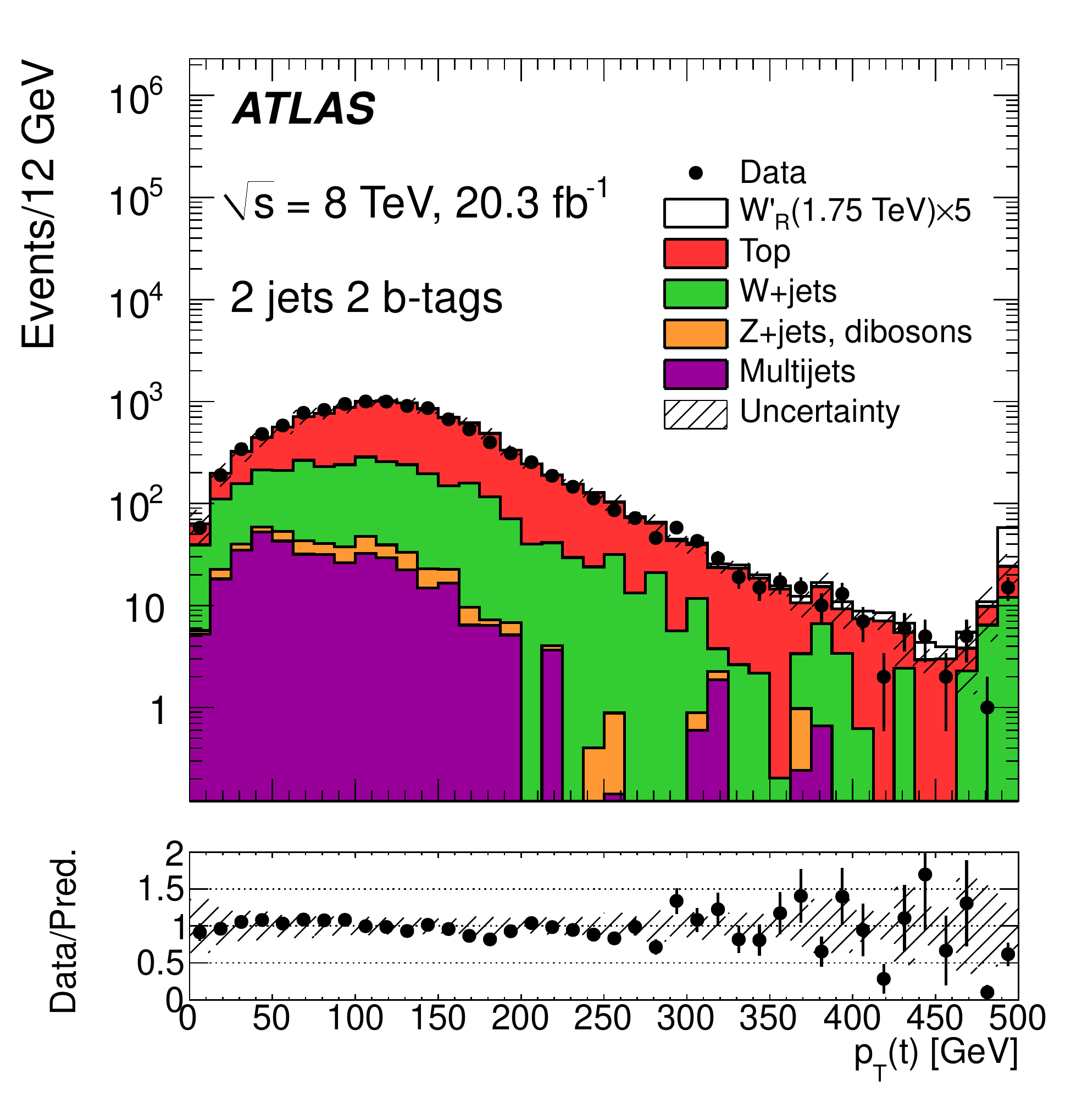}
\label{fig:pttop_SR_2jet}}
\subfigure[]{\includegraphics[width=0.40\textwidth]{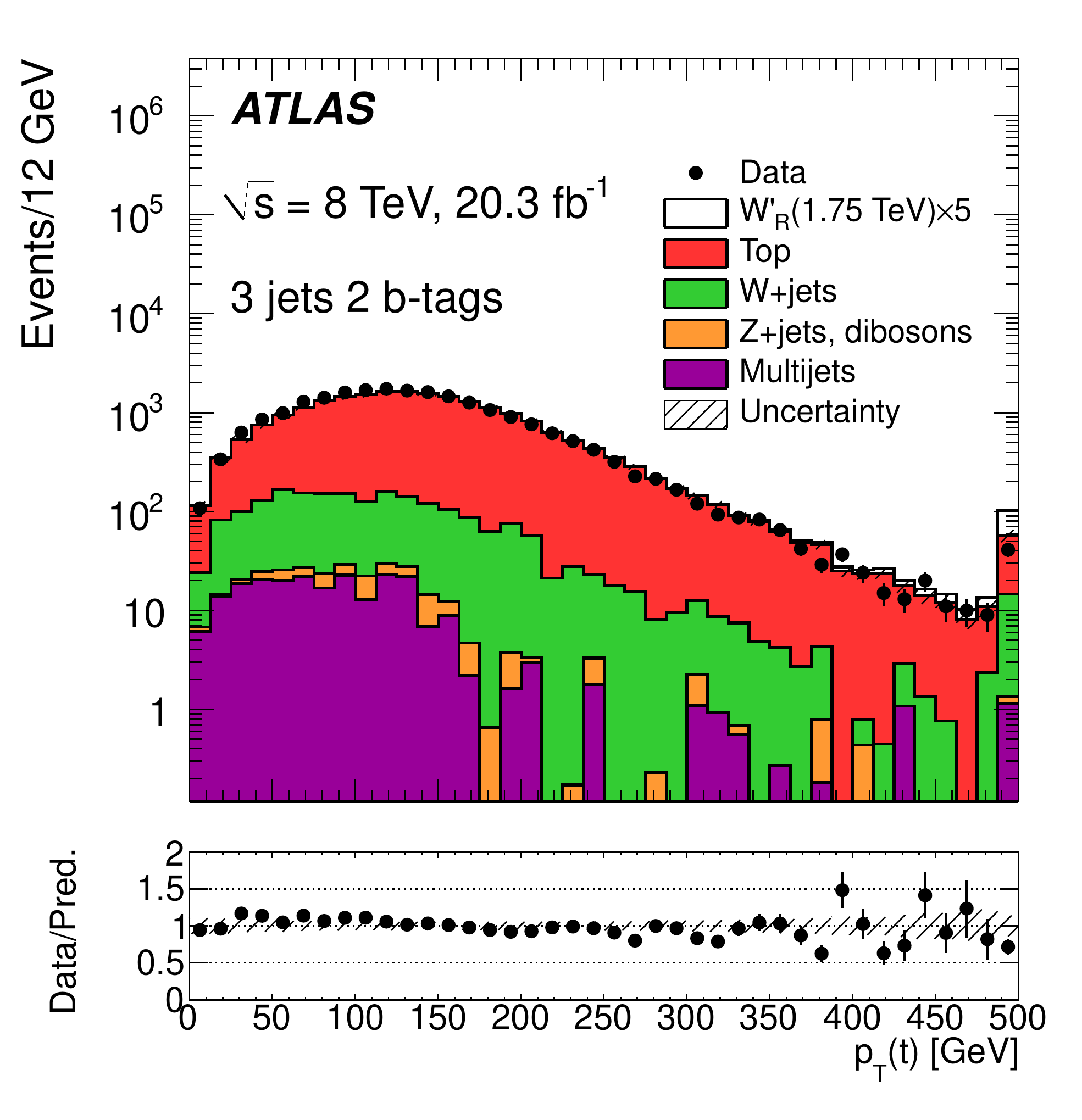}
\label{fig:pttop_SR_3jet}}
\caption{Distribution of, \subref{fig:mtb_SR_2jet} and~\subref{fig:mtb_SR_3jet}, the reconstructed invariant mass of the $t\bar{b}$ system
and, \subref{fig:pttop_SR_2jet} and~\subref{fig:pttop_SR_3jet}, of the reconstructed transverse momentum of the top quark, 
in the signal region for 2-jet and 3-jet events, respectively (electron and muon channels are combined).
The process labelled ``Top'' includes $t\bar{t}$ production and all three single top-quark production modes.
A signal contribution, amplified by a factor of five, corresponding to a $W'_\text{R}$ boson with a mass of 1.75~\tev\ is shown on top of the background distributions.
Uncertainty bands include normalisation uncertainties on all backgrounds and the uncertainty due to the limited size of the simulated samples.
The last histogram bin includes overflows.
(For interpretation of the references to colour in this figure legend, the reader is referred to the web version of this article.)
}
\label{fig:mtb_pttop_WpR1750}
\end{figure*}

A set of five variables is common to all four BDTs.
Two variables, \mtb\ and the transverse momentum of the reconstructed top quark, \pt$(t)$, provide the best separation power among all those considered
and are shown in Fig.~\ref{fig:mtb_pttop_WpR1750}.
The other three common variables are:
the angular separation\footnote{Defined as $\Delta R = \sqrt{(\Delta \phi)^2 + (\Delta \eta)^2}$.} 
between the jet associated with the $b$-jet originating from the $W'$ boson and the top-jet (denoted $b_t$), $\Delta R(b,b_t)$;
the transverse energy of the top-jet, $E_\text{T}(b_t)$,
and the aplanarity.\footnote{Aplanarity and sphericity are event shape variables
calculated from the sphericity tensor of the lepton and jet momenta~\cite{sphericity}.}

In addition, for the 2-jet channel, the following variables are used:
the angular separation between the top-jet and the $W$ boson, $\Delta R(b_t,W)$,
and the $\Delta \eta$ between the lepton and the top-jet, $\Delta \eta(\ell,b_t)$.
For the case of the right-handed $W'$-boson search the following variables are also used:
the sphericity;
the angular separation between the lepton and the $b$-jet originating from the $W'$ boson, $\Delta R(\ell,b)$;
the transverse momentum of the lepton, \pt$(\ell)$.
For the left-handed $W'$-boson search, three different variables are chosen:
the angle between the top-jet and the missing transverse momentum, $\Delta \phi(b_t,\MET)$;
the ratio of the transverse momenta of the top-jet and of the $b$-jet originating from the $W'$ boson, $\pt(b_t)/\pt(b)$,
and $m_\text{T}(W)$.

For events with three jets, the following variables are used in addition to the common set of variables:
$\Delta R(\ell,b_t)$;
the sphericity;
$\pt(b)$;
the invariant mass of the three jets $m(b,b_t,j)$.
Two more variables are used,
for the right-handed case only:
\pt$(\ell)$
and $\Delta R(b,W)$,
and for the left-handed case: 
$\Delta \phi(b_t,\MET)$
and 
$\pt(b_t)/\pt(b)$.

\begin{figure*}[!h!td]
\centering
\subfigure[]{\includegraphics[width=0.40\textwidth]{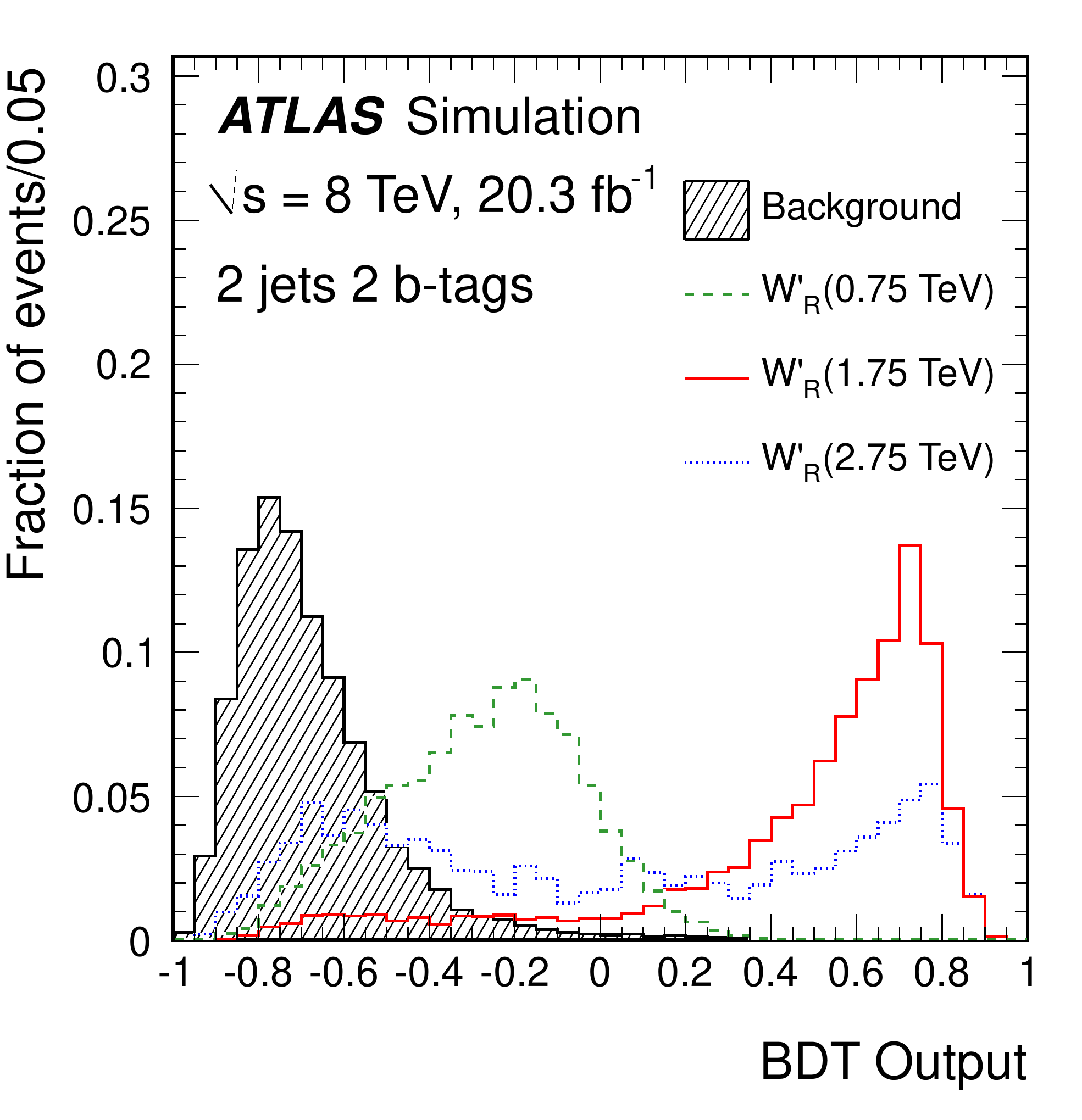}
\label{fig:BDTshapeR_2j}}
\subfigure[]{\includegraphics[width=0.40\textwidth]{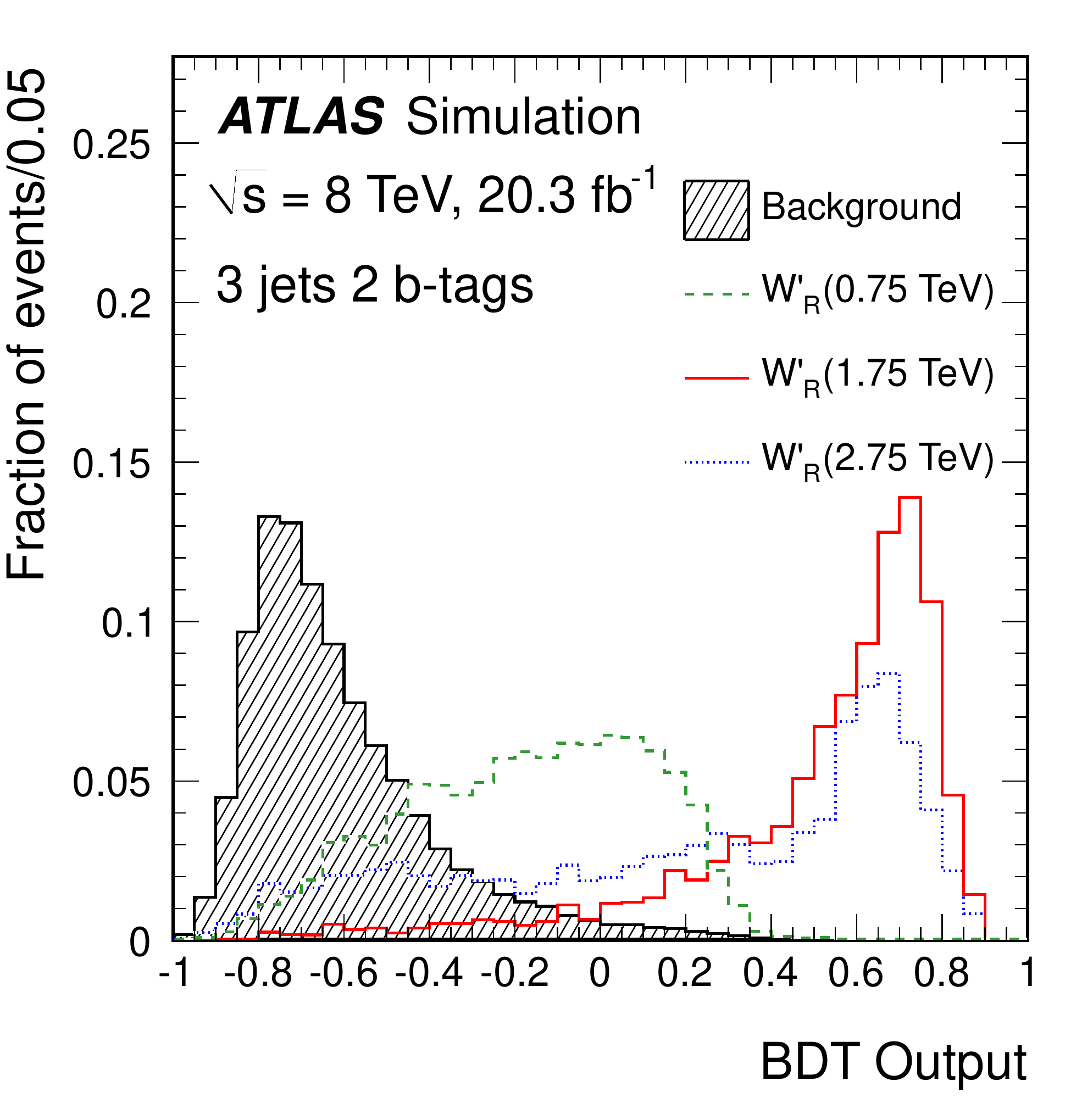}
\label{fig:BDTshapeR_3j}}
\caption{
Distributions of the BDT output values for the sum of all background processes (hatched histogram) and for 
three different mass values of the $W'_\text{R}$-boson signal (open histograms)
in \subref{fig:BDTshapeR_2j} the 2-jet and \subref{fig:BDTshapeR_3j} the 3-jet signal region.
Electron and muon channels are combined.
All distributions are normalised to unity.
} 
\label{fig:BDTshapeR}
\end{figure*}

Fig.~\ref{fig:BDTshapeR} shows the expected BDT output distributions, normalised to unity, in the signal region
for the electron and muon channels combined, for several simulated right-handed $W'$-boson samples and for the expected background.

%% file: Systematics.tex
\section{Systematic uncertainties}
\label{sec:systematics}

Systematic uncertainties can affect the shape and normalisation of the BDT output distributions.
They are split into the categories described below.

\noindent {\it Object modelling:} The main uncertainty in this category is due to 
uncertainties on $b$-tagging efficiency and mistagging rates~\cite{ATLAS-CONF-2014-004,ATLAS-CONF-2014-046}.
The resulting uncertainty on the event yield is 6\% for the total background contribution and 8--30\% for the signal.
The large uncertainties on the signal rates are due to additional
$b$-tagging uncertainties for jets with \pt\ above 300~\gev.
These uncertainties range from 3\% for $b$-jets with \pt\ of 50~\gev\ 
up to 15\% at 500~\gev\ and 35\% above.
The impact is sizeable for the signal where high-\pt\ jets stem from the $W'$-boson decay, 
in particular when the $W'$-boson mass is above 1~\tev.
The jet energy scale uncertainty depends on the \pt\ and $\eta$ of the reconstructed jet and includes the uncertainty on the $b$-jet energy scale. 
It results in an uncertainty on event yields of 1--6\% for the signal and 1--4\% for the background, depending on the channel.
The systematic uncertainty associated with the efficiency of the requirement on the jet vertex fraction results in rate variations of~2\%.
The impact of the jet energy resolution~\cite{JER} 
and the jet reconstruction efficiencies on signal and background rates is small. 
Uncertainties related to lepton energy scale and resolution as well as trigger and identification efficiencies
have a total effect of 2--4\% on the signal and background rates.
Another minor source of uncertainty comes from the propagation of the lepton and jet energy scale and resolution uncertainties to the \met.
The impact of pileup effects is negligible.

\noindent {\it Simulation modelling:} 
The dependence of the \ttbar\ event yield on additional radiation is evaluated by varying {\sc Pythia} parameters, 
while retaining consistency with a measurement of $t\bar{t}$ production with additional jet activity~\cite{ttbarveto}.
The variation in acceptance due to this source of uncertainty is 6--9\%.
The dependences of the \ttbar, single top-quark $s$-channel and $Wt$ event yields on the generator 
and parton showering simulation are estimated by comparing the nominal {\sc Powheg+Pythia} samples to samples produced 
using {\sc MC@NLO} v4.03~\cite{McAtNlo_a,McAtNlo_b} with the CT10 PDF set and interfaced to {\sc Herwig} 
and {\sc Jimmy} v4.31~\cite{Butterworth:1996zw}, for simulation of the underlying event and parton shower.
For the dominant \ttbar\ background only, the uncertainties arising from the choice of hadronisation and parton
shower models are also assessed by a comparison with {\sc Powheg+Herwig} samples. This comparison results in
a larger variation of the \ttbar\ event yields (6--11\%) and is thus taken as the associated systematic uncertainty.
For the $t$-channel single top-quark process, the comparison is performed between the nominal {\sc AcerMC+Pythia} sample
and a sample simulated with a{\sc MC@NLO} v2.1~\cite{aMCNLO} interfaced to {\sc Herwig}.
An uncertainty associated with the NLO calculation of $Wt$ production~\cite{WtProd} 
is evaluated by comparing the baseline sample generated with the diagram removal scheme
to a $Wt$ sample generated with the diagram subtraction scheme.
The PDF uncertainty on all signal and background simulated samples is estimated, following the {\sc PDF4LHC} recommendations~\cite{PDF4LHC},
by taking the envelope of {\sc CT10}, {\sc MWST2008nlo68cl}~\cite{MSTW} and 
{\sc NNPDF23}~\cite{NNPDF} PDF sets uncertainties at 68\% CL and 
normalising to the nominal cross-section.
The uncertainties on event yields are~9\% for signal and 3--5\% for the total background.
The statistical uncertainty due to the limited size of the simulated samples is also taken into account.

\noindent {\it Background normalisation:} Theoretical uncertainties on cross-sections are 
--5.9/+5.1\% for the $t\bar{t}$ process, 
$-2.1$/+3.9\% and 3.9\% for single top-quark production in the $t$-channel and $s$-channel respectively, 
and 6.8\% for the $Wt$ channel process.
For the $W$+jets background in the 2-jet channel an average total uncertainty of 50\% is used 
as the result of the propagation, in the data-driven method described in Section~\ref{sec:bgestimation}, of the following uncertainties:
theoretical uncertainties on \ttbar, single top-quark and $Z$+jets/diboson cross-sections, modelling uncertainties of the \ttbar\ process,
uncertainty on the multijet rate, and systematic uncertainties on the jet energy scale and $b$-tagging efficiency.
The theoretical normalisation uncertainty used in the 3-jet channel is 42\%. This estimate is derived from
the uncertainty on the inclusive cross-section of $W$-boson production~\cite{FEWZ} (4\%) and the uncertainty
on the cross-section ratios of $W$-boson production associated with $n+1$ jets to $W$-boson production associated 
with $n$ jets~\cite{Berends} (24\% per jet, added in quadrature).
An uncertainty of 42\% is conservatively assigned to the diboson and $Z$+jets rates, which represent very small backgrounds.
A systematic uncertainty of 50\% on the rate of the multijet background is estimated from a study of
uncertainties on the efficiencies and fake rates.

\noindent {\it Luminosity:} The uncertainty on the integrated luminosity is 2.8\%. 
It is derived, following the same methodology as that detailed in Ref.~\cite{lumiref}, from a preliminary calibration of the luminosity scale 
derived from beam-separation scans performed in November 2012.

The impact on the signal and background 
event yields of the main object and modelling uncertainties is summarised in Table~\ref{tab:systs}.

\begin{table}[!h!tdbp]
    \caption{\label{tab:systs} 
Impact of the main sources of object and modelling systematic uncertainties on the signal and background event yields 
(only yield differences above 1\% are shown). 
The quoted modelling uncertainty is on $t\bar{t}$ yields only.
}
    \begin{center}\small
      \begin{tabular}{l|cc}
              \hline 
Source                  & $W'_\text{R}$ (1.75 \tev) & Background  \\
\hline                                               
$b$-tagging efficiency  & 27\%	                   & 6\%         \\
Jets	  		& 4\%	                   & 1--4\%      \\
Lepton                  & 2\%	                   & 2--4\%      \\
$t\bar{t}$ modelling    & 		           & 8--14\% 	 \\
PDF			& 9\%		           & 3--5\%      \\
\hline
\end{tabular}       
\end{center}  
\end{table}

%% file: Result.tex
\section{Results}
\label{sec:result}

\begin{figure*}[hbtp]
\centering
\subfigure[]{\includegraphics[width=0.40\textwidth]{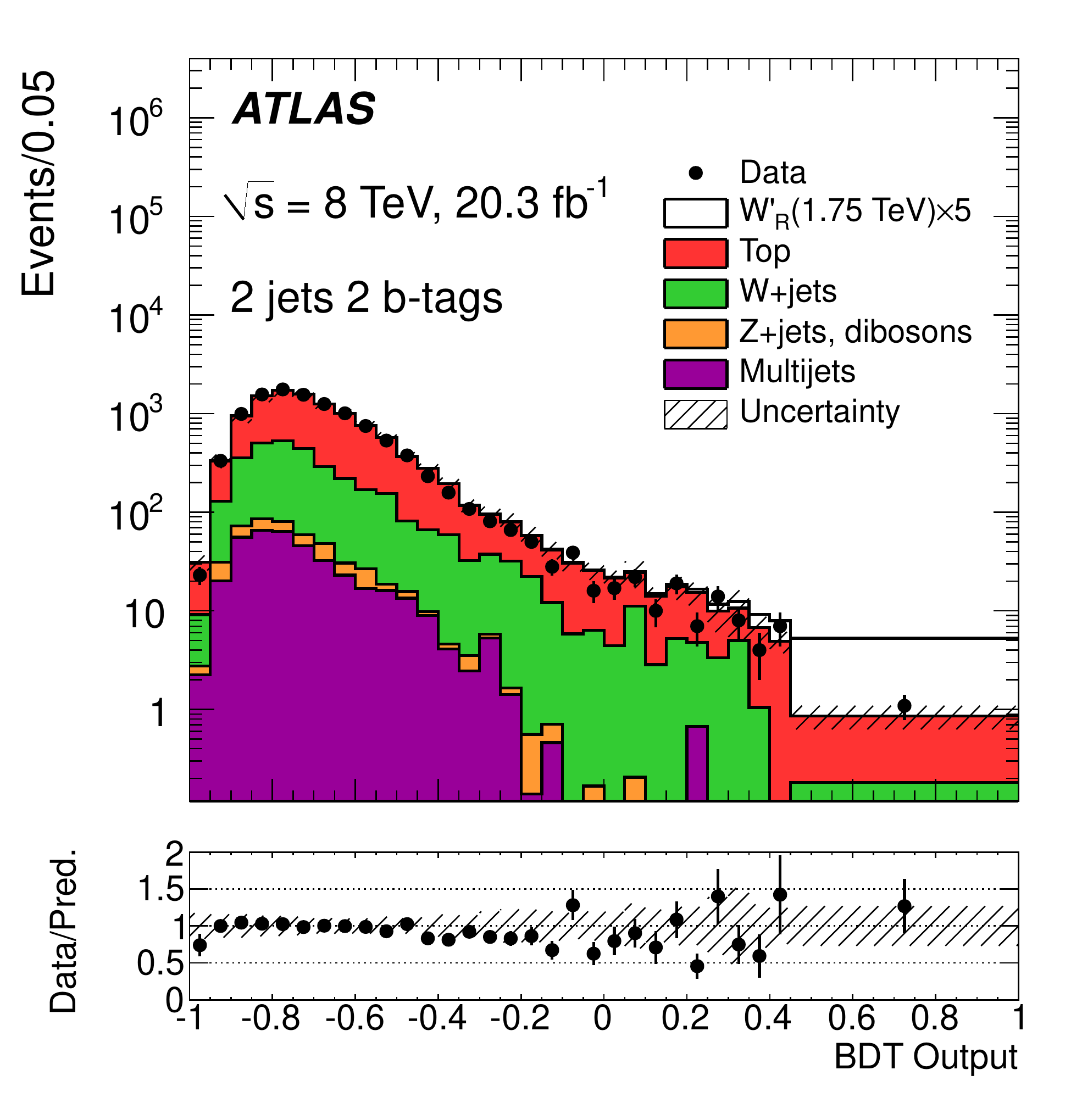}
\label{fig:BDT_SR_2jet}}
\subfigure[]{\includegraphics[width=0.40\textwidth]{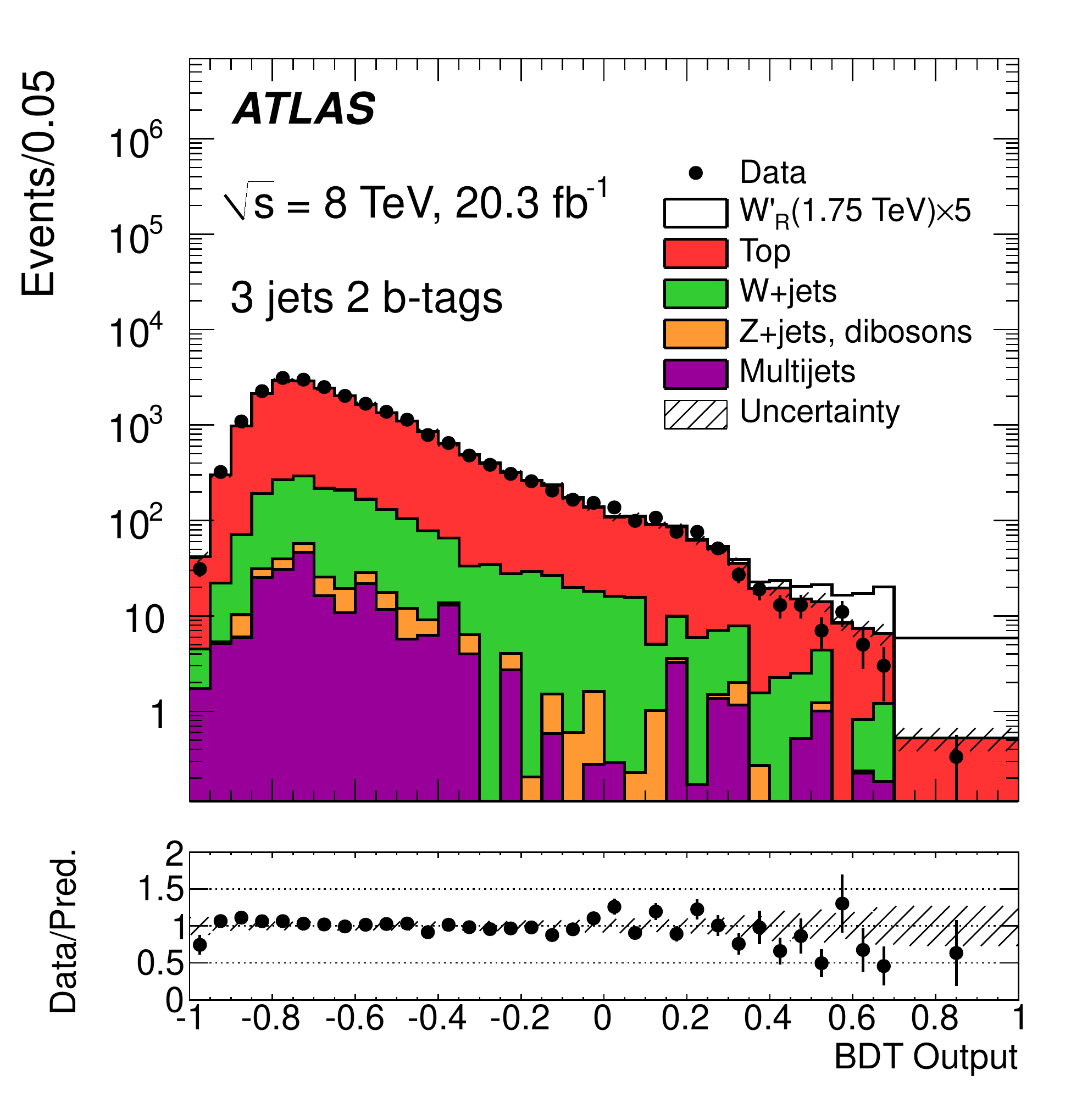}
\label{fig:BDT_SR_3jet}}
\caption{BDT output distributions in the signal region, in \subref{fig:BDT_SR_2jet} 2-jet and \subref{fig:BDT_SR_3jet} 3-jet events 
(electron and muon channels are combined).
The process labelled ``Top'' includes $t\bar{t}$ production and all three single top-quark production modes.
A signal contribution, amplified by a factor of five, corresponding to a $W'_\text{R}$ boson with a mass of 1.75~\tev\ is shown on top of the background distributions.
The uncertainty band includes normalisation uncertainties on all backgrounds and the uncertainty due to the limited size of the simulated samples.
(For interpretation of the references to colour in this figure legend, the reader is referred to the web version of this article.)
}
\label{fig:SignalBDT}
\end{figure*}

Table~\ref{tab:SR2j2t} reports the numbers of data events and expected signal and background events for an integrated luminosity of \lumi\ 
in the signal region for 2-jet and 3-jet events, where the electron and muon channels are combined.
Fig.~\ref{fig:SignalBDT} shows the BDT output distributions in the signal region.
The signal contribution corresponding to a $W'_\text{R}$ boson with a mass of 1.75~\tev\ is shown, amplified by a factor of five,
on top of the background distributions.

\begin{table}[hbtp]\small
\begin{center}
\caption{\label{tab:SR2j2t} 
The numbers of expected signal and background events and the numbers of observed data events are shown in the 2-jet and 3-jet signal regions.
The quoted uncertainties account for all systematic effects as well as for the statistical uncertainty due to 
the limited number of events in the simulated samples.}
\begin{tabular}{l r@{ $\pm$ }l@{ } c r@{ $\pm$ }l@{ }}
\hline
 & \multicolumn{2}{c}{2-jet 2-tag} & & \multicolumn{2}{c}{3-jet 2-tag}\\
\hline 
          $W'_\text{R}$ (0.5 \tev) &    15400 &     1600 &&     9950 &     1100 \\
          $W'_\text{R}$ (1.0 \tev) &      720 &      140 &&      800 &      140 \\
          $W'_\text{R}$ (1.5 \tev) &       49 &       15 &&       67 &       17 \\
          $W'_\text{R}$ (2.0 \tev) &      4.9 &      1.4 &&      7.3 &      2.0 \\
          $W'_\text{R}$ (2.5 \tev) &     0.8  &      0.2 &&     1.0  &      0.3 \\
          $W'_\text{R}$ (3.0 \tev) &     0.26 &     0.05 &&     0.29 &     0.06 \\
\hline
                $t\bar{t}$  &     6450 &     1100 &&    17700 &     2500 \\
     Single-top $t$-channel &      900 &      360 &&     1190 &      230 \\
            Single-top $Wt$ &      320 &       50 &&      850 &      210 \\
     Single-top $s$-channel &      250 &       30 &&      137 &       20 \\
                  $W$+jets  &     2700 &     1300 &&     1800 &      900 \\
                   Diboson  &      100 &       50 &&       70 &       30 \\
                  $Z$+jets  &       17 &        7 &&       14 &        6 \\
                  Multijets &      380 &      190 &&      210 &      105 \\
\hline
           Total background &    11100 &     1900 &&    22000 &     3100 \\
\hline 
Data & \multicolumn{2}{l}{11039} & & \multicolumn{2}{l}{22555}\\
\hline
\end{tabular}       
\end{center}  
\end{table}

%% file: Limit_Setting.tex
No excess in data is observed over the full BDT output distributions.
Therefore, the BDT distributions in the 2-jet and 3-jet channels, where electron and muon samples are separated,
are used in a combined statistical analysis to calculate exclusion limits on the production cross-section of 
the left-handed or right-handed $W'$ boson as a function of its mass.
The case of left-handed $W'$-boson production is treated in two different ways. 
In the first, the interference between $W'_\text{L}$-boson and SM $s$-channel single top-quark production is neglected.
The limits obtained are then valid only for the left-handed signal without the interference contribution.
In the second, the interference effect is accounted for by considering 
the $pp \to W'_\text{L}/W \to t\bar{b}$ process as a unique signal and by setting limits on the 
cross-section of $W'_\text{L}/W \to t\bar{b}$ production, as a function of the $W'_\text{L}$-boson mass.

Hypothesis testing is performed with the CL$_s$ procedure~\cite{MCLimit,CLs} using a log-likelihood ratio as the test statistic,
defined as the ratio of two hypotheses: the test hypothesis, which admits the presence of a $W'$-boson signal in addition to the SM 
backgrounds, and the null hypothesis, which considers only SM backgrounds. 
For a given hypothesis, the combined likelihood is the product of the likelihoods for the four individual channels considered
(2/3-jet and electron/muon samples), each of which is a product of Poisson probabilities over the bins of the BDT output histogram. 
Pseudo-experiments are generated for both hypotheses, taking into account per-bin statistical fluctuations of the total predictions
according to Poisson statistics, as well as Gaussian fluctuations describing the effect of systematic uncertainties.
Correlations across bins, channels, and processes are taken into account.
In order to reduce the impact of systematic uncertainties on the sensitivity of the search, a nuisance parameter
corresponding to a scaling factor on the overall $t\bar{t}$ yield is fitted to data during the statistical analysis.
This scaling factor is found to be consistent with unity.

\begin{figure*}[hbtp]
\centering
\subfigure[]{\includegraphics[width=0.40\textwidth]{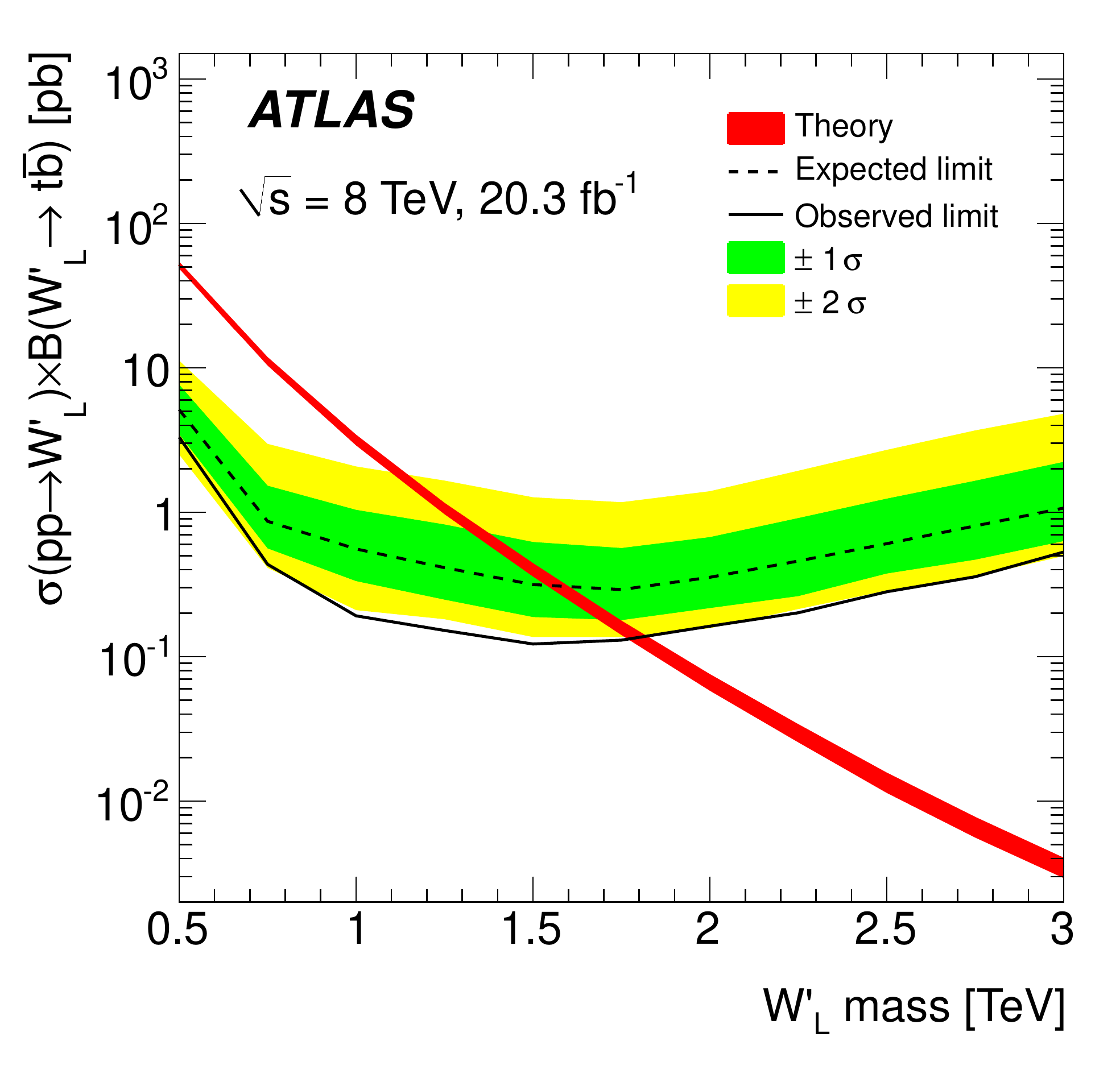}
\label{fig:Limit_left}}
\subfigure[]{\includegraphics[width=0.40\textwidth]{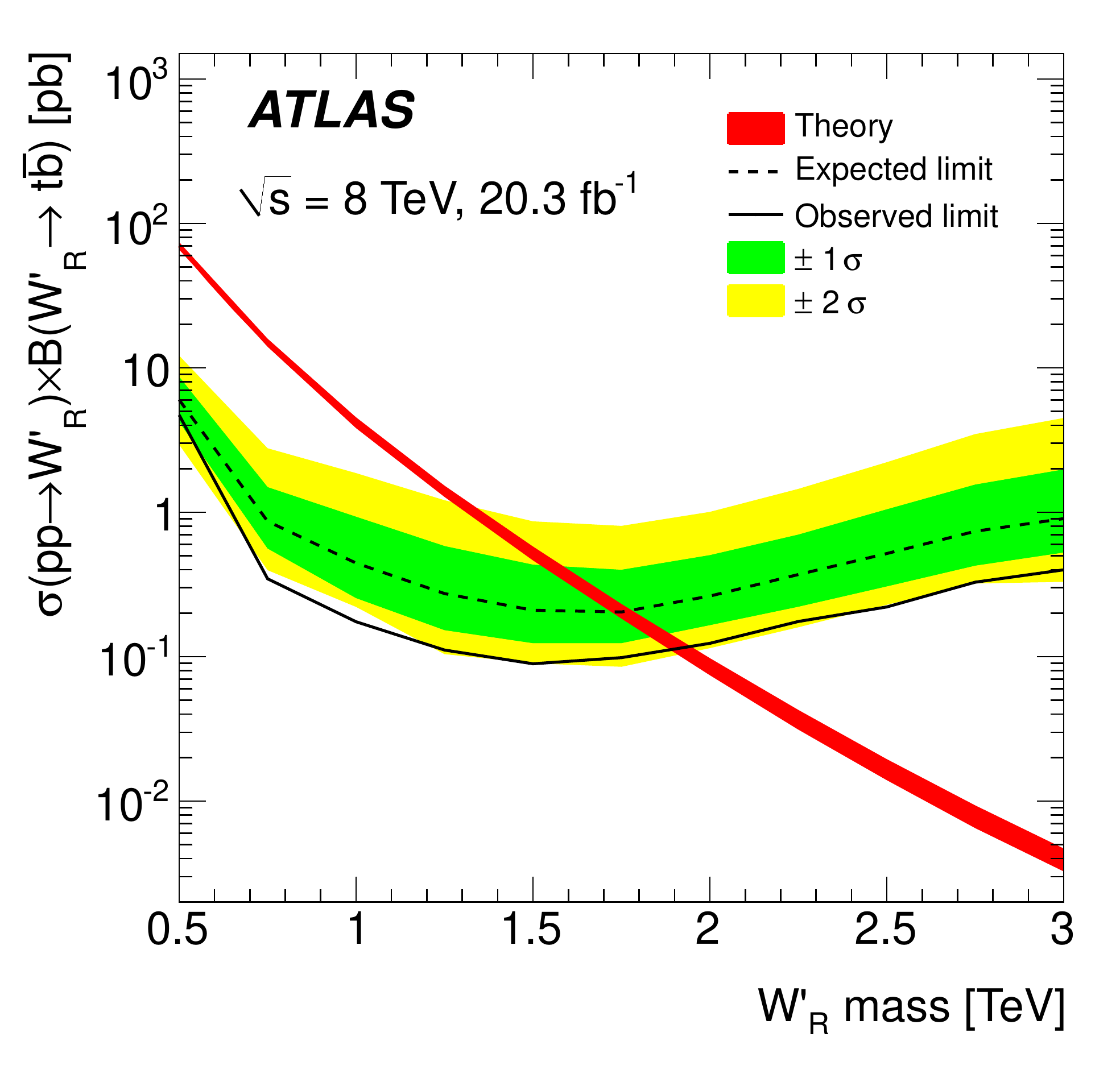}
\label{fig:Limit_right}}
\caption{Observed and expected 95\% CL limits on the $W'$-boson cross-section times branching ratio, as a function of the $W'$-boson mass,
for \subref{fig:Limit_left} left-handed and \subref{fig:Limit_right} right-handed $W'$ bosons. 
Theoretical predictions of the signal cross-sections~\cite{Duffty:2012rf}
(where leptonic decay of the $W'_\text{R}$ boson is not allowed and interference of the $W'_\text{L}$ boson with the $s$-channel single top-quark production is not considered)
are represented by a solid red line.
Theoretical uncertainties, shown as a band, range from about 5\% for small $W'$-boson masses to 20\% for large masses
and are dominated by the uncertainty from the {\sc CTEQ6.6}~\cite{cteq6} NLO PDFs.
(For interpretation of the references to colour in this figure legend, the reader is referred to the web version of this article.)
}
\label{fig:LimitBDTXSsyst}
\end{figure*}

Fig.~\ref{fig:LimitBDTXSsyst} shows the observed and expected 95\% CL limit on the $W'$-boson cross-section times branching ratio, 
as a function of the $W'$-boson mass, for left-handed (without the interference term) and right-handed  $W'$-boson couplings.
The fact that the observed limits are lower than expected can be explained by a deficit in data, compared to the predicted 
number of background events, in the high BDT output region. This deficit is also visible in the tails of the \mtb\ distributions: 
it is rather localised at 1.1~\tev\ in 2-jet events, and widespread in 3-jet events (see Fig.~\ref{fig:mtb_pttop_WpR1750}).
In the BDT distributions, however, this deficit is seen mostly at BDT output values higher than~0 in both 2-jet and 3-jet channels, 
in a region where the $W'$-boson distributions peak (for $W'$-boson masses of 1~\tev\ and above).  In addition, because of the rather 
large width of the signal BDT distributions, all $W'$-boson cross-section limits are affected by these fluctuations in the data.

The point where the measured cross-section limit crosses the theory curve defines the 95\% CL lower limit on the $W'$-boson mass.
Masses below 1.92 (1.80, 1.70)~\tev\ are excluded for right-handed (left-handed without and with interference) $W'$ bosons, 
while the expected limit is 1.75 (1.57, 1.54)~\tev.
The theoretical cross-section times branching ratio values and the observed limits are reported 
in Table~\ref{tab:Limits} for several left-handed and right-handed $W'$-boson hypotheses.

\begin{table*}[hbt]
    \caption{\label{tab:Limits} 
Theoretical cross-section times branching ratio values and observed 95\% CL limits 
for left-handed and right-handed $W' \to t\bar{b}$ production (columns two to five)
and for $W'_\text{L}/W \to t\bar{b}$ production, including the interference term (last two columns).
In the latter case, results are not shown for $W'$-boson masses above 2~\tev\ due to the lack of
$W'$ events in the $W'_\text{L}/W \to t\bar{b}$ simulated samples generated at high values of the $W'$-boson mass.}
    \begin{center}\small
      \begin{tabular}{c|cc|cc|cc}
              \hline 
      &  \multicolumn{2}{c|}{$W'_\text{L} \to t\bar{b}$} & \multicolumn{2}{c|}{$W'_\text{R} \to t\bar{b}$} & \multicolumn{2}{c}{$W'_\text{L}/W \to t\bar{b}$} \\
$W'$ mass &  Theory & Obs. limit & Theory & Obs. limit & Theory & Obs. limit \\
$[$\tev] & [pb] & [pb] & [pb] & [pb] & [pb] & [pb] \\
\hline
0.5      &  52	     &  3.3	& 70	& 4.7   & 53	& 3.5  \\
1.0      &  3.2	     &  0.19	& 4.2	& 0.17	& 7.6	& 0.51 \\
1.5      &  0.40     &	0.12	& 0.52	& 0.089	& 5.5	& 2.0  \\	 
2.0      &  0.067    &	0.16	& 0.086	& 0.12	& 5.4	& 19   \\	
2.5      &  0.014    &	0.28	& 0.017	& 0.22	& -	& -    \\
3.0      &  0.0035   &	0.53	& 0.004	& 0.40  & -	& -    \\
              \hline 
            \end{tabular}       
          \end{center}  
  \end{table*}

\begin{figure*}[p]
\centering
\subfigure[]{\includegraphics[width=0.40\textwidth]{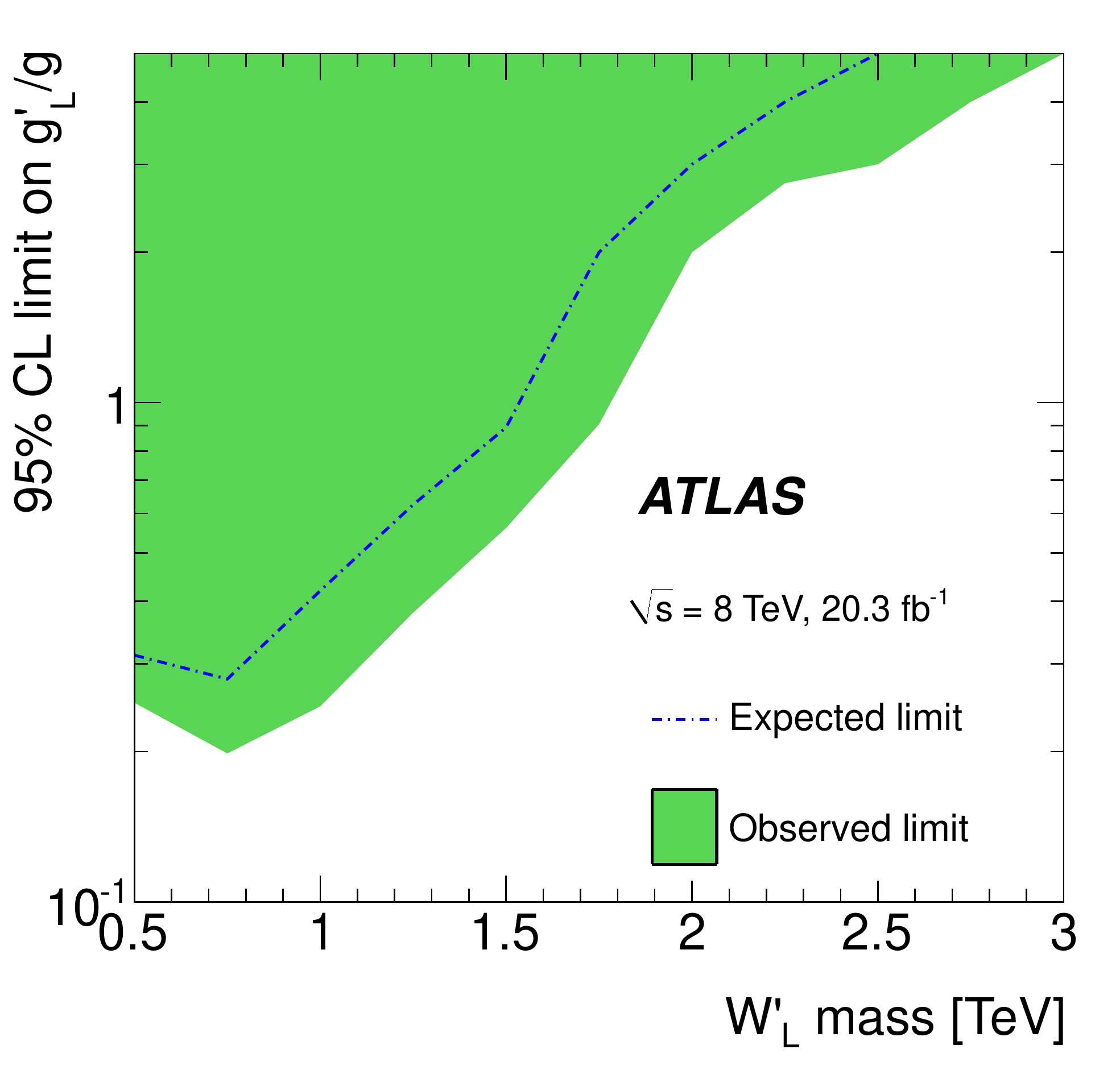}
\label{fig:gprime_left}}
\subfigure[]{\includegraphics[width=0.40\textwidth]{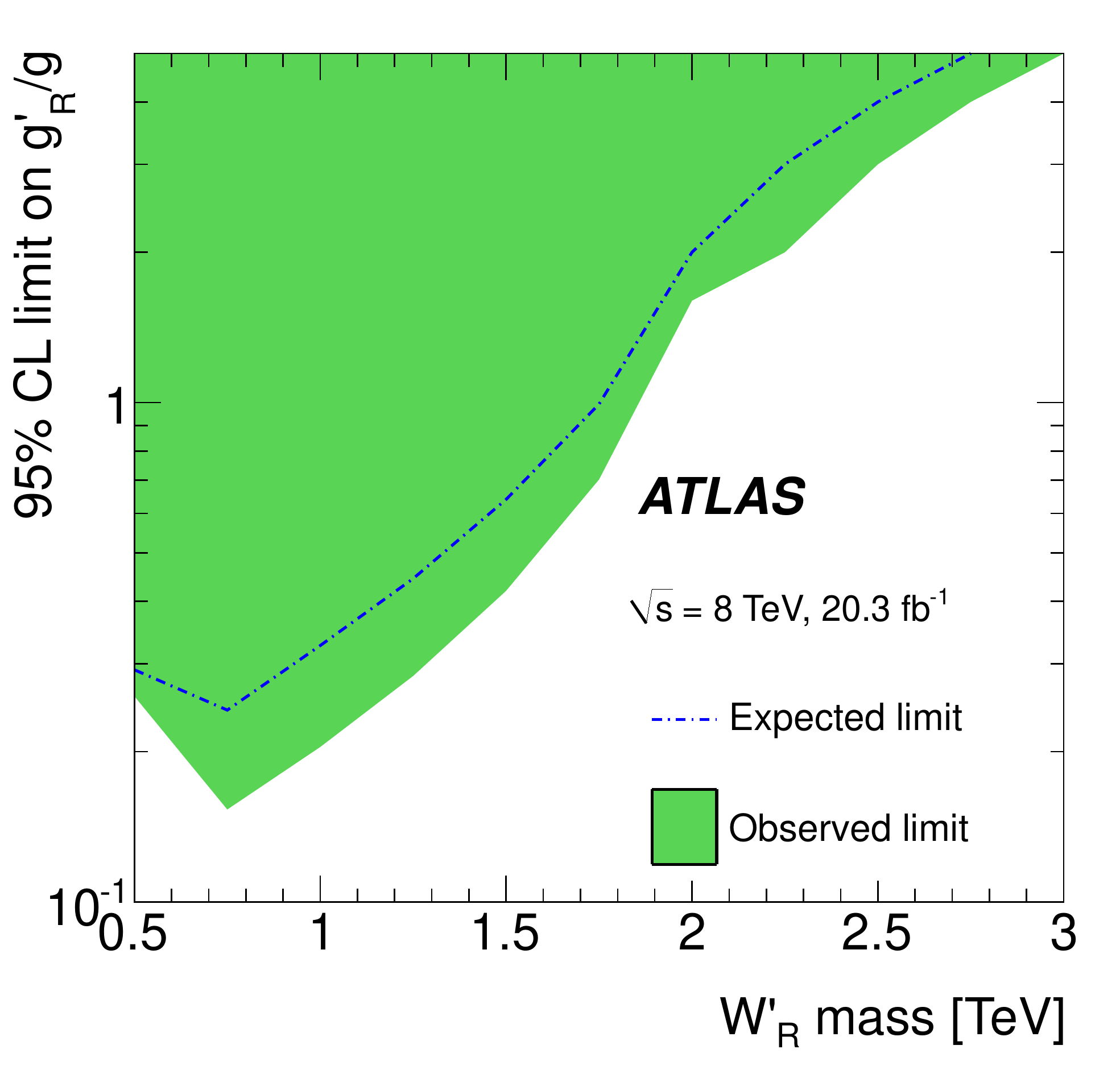}
\label{fig:gprime_right}}
\caption{Observed and expected regions, on the $g'/g$ vs mass of the $W'$-boson plane, that are excluded at 95\% CL,
for~\subref{fig:gprime_left} left-handed (no interference) and~\subref{fig:gprime_right} right-handed $W'$ bosons. 
}
\label{fig:LimitBDTGsyst}
\end{figure*}

\begin{figure*}[p]
\centering
\subfigure[]{\includegraphics[width=0.40\textwidth]{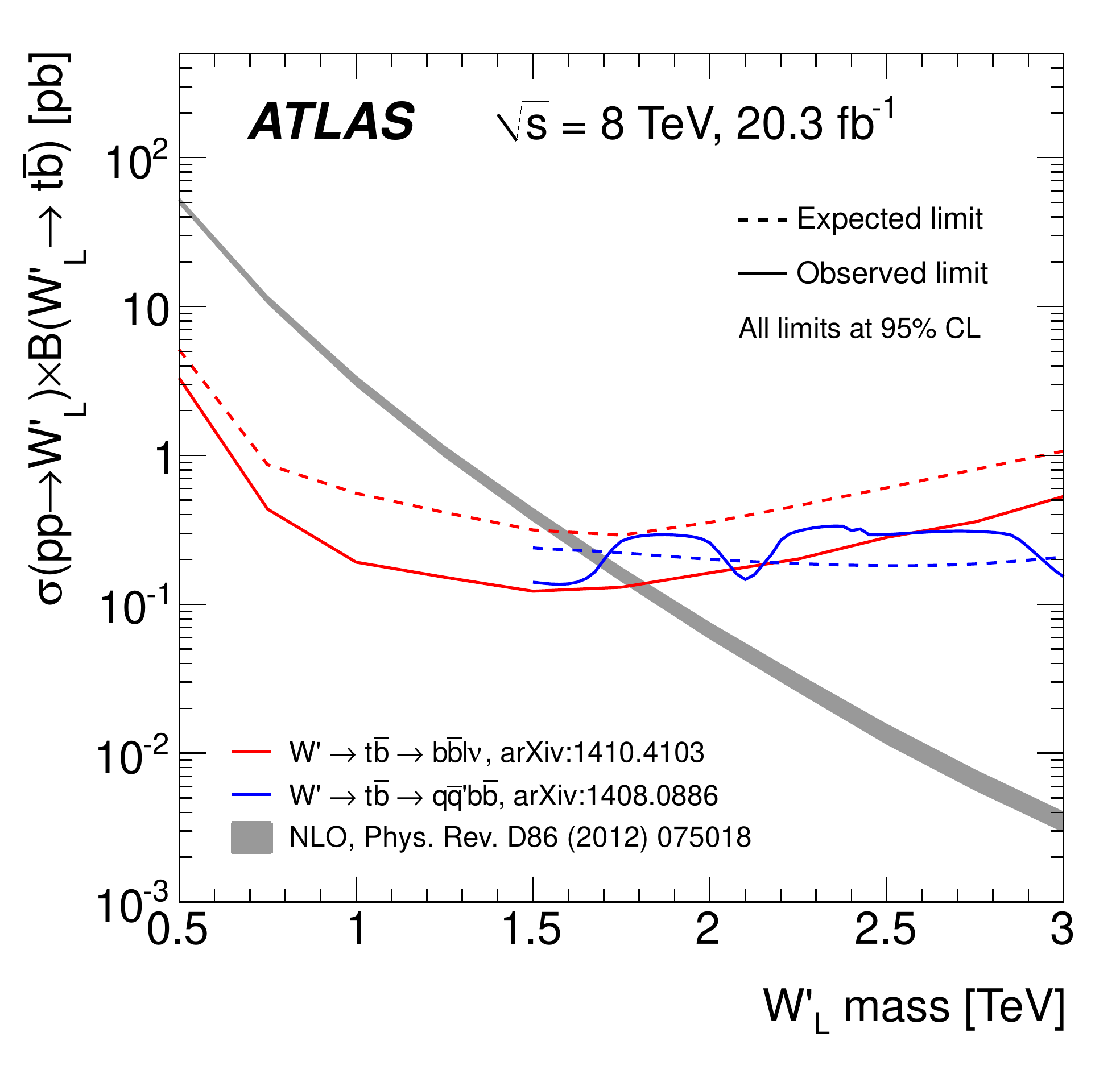}
\label{fig:Limit_left_summary}}
\subfigure[]{\includegraphics[width=0.40\textwidth]{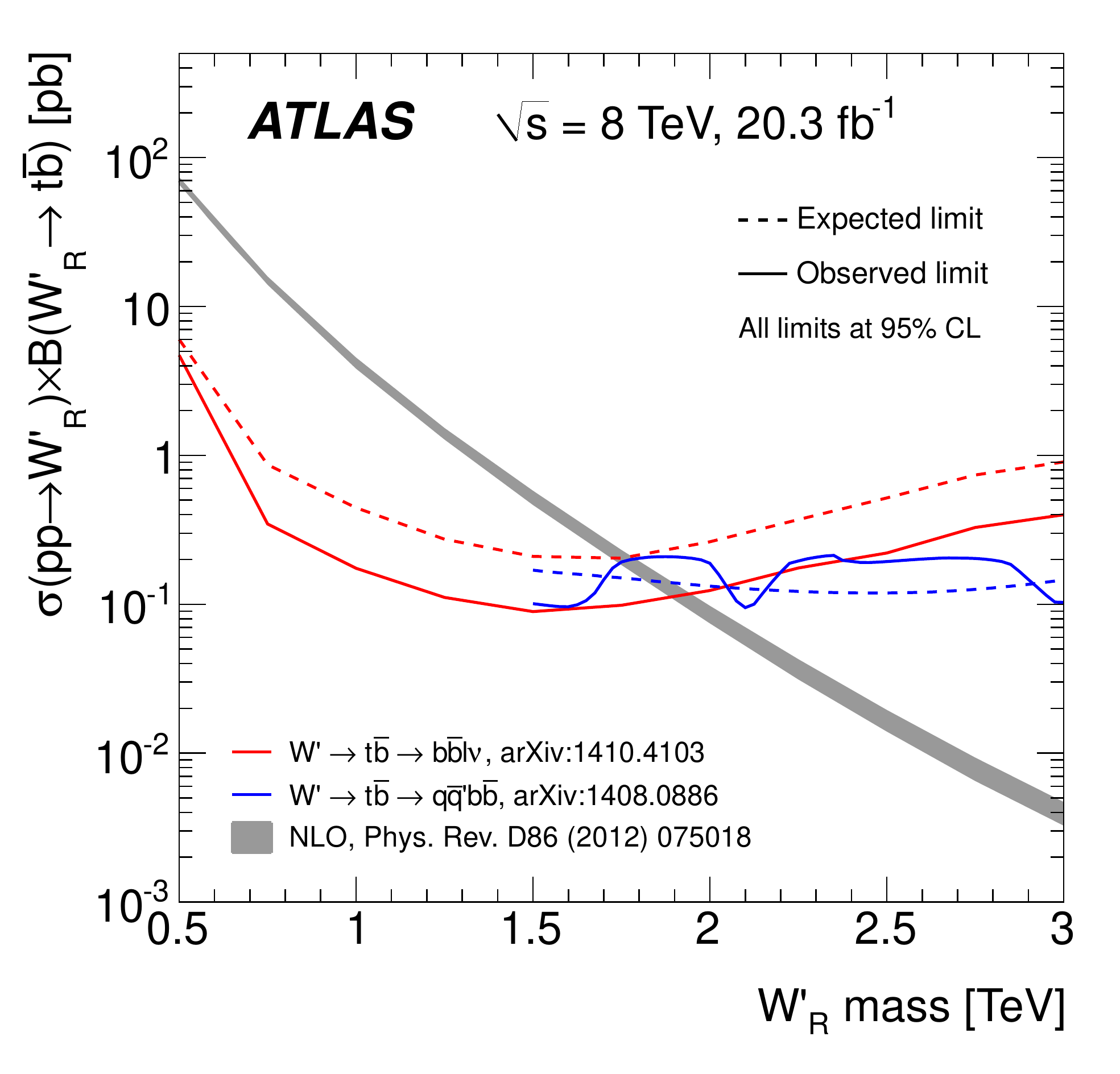}
\label{fig:Limit_right_summary}}
\caption{Observed and expected 95\% CL limits on the $W'$-boson cross-section times branching ratio, as a function of the $W'$-boson mass,
for \subref{fig:Limit_left_summary} left-handed and \subref{fig:Limit_right_summary} right-handed $W'$ bosons. 
Results are shown for the present analysis (red lines) together with the limits obtained by a search for $W' \to t\bar{b}$ 
boson production in the fully hadronic channel~\cite{ATLASWhad} (blue lines).
Theoretical predictions of the signal cross-sections~\cite{Duffty:2012rf} are represented by a solid grey line.
(For interpretation of the references to colour in this figure legend, the reader is referred to the web version of this article.)
}
\label{fig:XSSummaryPlot}
\end{figure*}

Limits on the ratio of couplings $g'/g$ as a function of the $W'$-boson mass can be derived from the limits on the $W'$-boson cross-section.
Limits can also be set for $g'/g>1$, as models remain perturbative up to a ratio of about five~\cite{Duffty:2012rf}.
A given hypothesis $g'$ for a $W'$ boson of mass $m_{W'}$ is excluded if the resulting theoretical cross-section is higher than the cross-section
limit derived previously.
The $W'$-boson cross-section has a non-trivial dependence on the coupling $g'$, coming from the variation of the resonance width, 
which is proportional to $g'^2$. 
The scaling of the $W'$-boson cross-section as a function of $g'/g$ and $m_{W'}$ is estimated using {\sc MadGraph}.
The impact of NLO corrections on this scaling is found to be at most a few percent and is neglected.
In addition, specific signal samples (see Section~\ref{sec:samples})
are used in order to take into account the effect on the acceptance and on kinematical distributions
of the increased signal width (compared to the nominal samples) for values of $g'/g>1$.
Fig.~\ref{fig:LimitBDTGsyst} shows the observed and expected 95\% CL limits on the ratio $g'/g$,
as a function of $m_{W'}$, for left-handed (no interference) and right-handed $W'$-boson couplings.
The lowest observed (expected) limits on $g'/g$, obtained for a $W'$-boson mass of 0.75~\tev, are 0.20 (0.28) and 0.16 (0.24) for $W'_\text{L}$ and $W'_\text{R}$, respectively.

Figure~\ref{fig:XSSummaryPlot} shows the $W'$-boson cross-section limits of Fig.~\ref{fig:LimitBDTXSsyst} together with the limits
obtained by a search for $W' \to t\bar{b}$ boson production in the fully hadronic channel~\cite{ATLASWhad}
performed at $\sqrt{s} = 8$~\tev\ with the ATLAS detector.

%% file: Acknowledgements.tex

\section*{Acknowledgements}

We thank CERN for the very successful operation of the LHC, as well as the
support staff from our institutions without whom ATLAS could not be
operated efficiently.

We acknowledge the support of ANPCyT, Argentina; YerPhI, Armenia; ARC,
Australia; BMWFW and FWF, Austria; ANAS, Azerbaijan; SSTC, Belarus; CNPq and FAPESP,
Brazil; NSERC, NRC and CFI, Canada; CERN; CONICYT, Chile; CAS, MOST and NSFC,
China; COLCIENCIAS, Colombia; MSMT CR, MPO CR and VSC CR, Czech Republic;
DNRF, DNSRC and Lundbeck Foundation, Denmark; EPLANET, ERC and NSRF, European Union;
IN2P3-CNRS, CEA-DSM/IRFU, France; GNSF, Georgia; BMBF, DFG, HGF, MPG and AvH
Foundation, Germany; GSRT and NSRF, Greece; RGC, Hong Kong SAR, China; ISF, MINERVA, GIF, I-CORE and Benoziyo Center, Israel; INFN, Italy; MEXT and JSPS, Japan; CNRST, Morocco; FOM and NWO, Netherlands; BRF and RCN, Norway; MNiSW and NCN, Poland; GRICES and FCT, Portugal; MNE/IFA, Romania; MES of Russia and ROSATOM, Russian Federation; JINR; MSTD,
Serbia; MSSR, Slovakia; ARRS and MIZ\v{S}, Slovenia; DST/NRF, South Africa;
MINECO, Spain; SRC and Wallenberg Foundation, Sweden; SER, SNSF and Cantons of
Bern and Geneva, Switzerland; NSC, Taiwan; TAEK, Turkey; STFC, the Royal
Society and Leverhulme Trust, United Kingdom; DOE and NSF, United States of
America.

The crucial computing support from all WLCG partners is acknowledged
gratefully, in particular from CERN and the ATLAS Tier-1 facilities at
TRIUMF (Canada), NDGF (Denmark, Norway, Sweden), CC-IN2P3 (France),
KIT/GridKA (Germany), INFN-CNAF (Italy), NL-T1 (Netherlands), PIC (Spain),
ASGC (Taiwan), RAL (UK) and BNL (USA) and in the Tier-2 facilities
worldwide.

%% file: atlas_authlist.tex
\begin{flushleft}
{\Large The ATLAS Collaboration}

\bigskip

G.~Aad$^{\rm 85}$,
B.~Abbott$^{\rm 113}$,
J.~Abdallah$^{\rm 152}$,
S.~Abdel~Khalek$^{\rm 117}$,
O.~Abdinov$^{\rm 11}$,
R.~Aben$^{\rm 107}$,
B.~Abi$^{\rm 114}$,
M.~Abolins$^{\rm 90}$,
O.S.~AbouZeid$^{\rm 159}$,
H.~Abramowicz$^{\rm 154}$
% ...
\bigskip
\\
% ...
$^{11}$ Institute of Physics, Azerbaijan Academy of Sciences, Baku, Azerbaijan\\
% ...
$^{85}$ CPPM, Aix-Marseille Universit{\'e} and CNRS/IN2P3, Marseille, France\\
% ...
$^{90}$ Department of Physics and Astronomy, Michigan State University, East Lansing MI, United States of America\\
% ...
$^{107}$ Nikhef National Institute for Subatomic Physics and University of Amsterdam, Amsterdam, Netherlands\\
% ...
$^{113}$ Homer L. Dodge Department of Physics and Astronomy, University of Oklahoma, Norman OK, United States of America\\
$^{114}$ Department of Physics, Oklahoma State University, Stillwater OK, United States of America\\
% ...
$^{117}$ LAL, Universit{\'e} Paris-Sud and CNRS/IN2P3, Orsay, France\\
% ...
$^{152}$ Institute of Physics, Academia Sinica, Taipei, Taiwan\\
% ...
$^{154}$ Raymond and Beverly Sackler School of Physics and Astronomy, Tel Aviv University, Tel Aviv, Israel\\
% ...
$^{159}$ Department of Physics, University of Toronto, Toronto ON, Canada\\
% ...
\end{flushleft}
